\documentclass[12 pt]{article}

\usepackage[dvips]{graphicx}
\usepackage{amsfonts}

\AtBeginDocument{
  \pretolerance-1
  \tolerance1900
  \adjdemerits10000
  \emergencystretch10mm   
  \clubpenalty10000       
  \widowpenalty10000
  \displaywidowpenalty10000
  }

\begin{document}
\begin{center}

{\bf\Large Multiplication of Qubits in a Doubly Resonant Bichromatic Field}

{A.\,P.\,Saiko$^{*}$\/\thanks{e-mail: saiko@ifttp.bas-net.by},
 R. Fedaruk$^{+}$}
\end{center}

\begin{abstract}
\footnotesize  Multiplication of spin qubits arises at double
resonance in a bichromatic field when the frequency of the
radio-frequency (rf) field is close to that of the Rabi
oscillation in the microwave field, provided its frequency equals
the Larmor frequency of the initial qubit. We show that the
operational multiphoton transitions of dressed qubits can be
selected by the choice of both the rotating frame and the rf
phase. In order to enhance the precision of dressed qubit
operations in the strong-field regime, the counter-rotating
component of the rf field is taken into account.
\end{abstract}
{\footnotesize {\bf PACS}: 03.67.Pp, 33.40.+f, 33.35.+r}

Theoretical models of quantum computations assume the existence of
an ideal two-level quantum system (qubit) and the possibility of
an exact description of the qubit's interaction with external
electromagnetic fields~[1]. It is known that the resonant
interaction between electromagnetic radiation and qubit induces
Rabi oscillations, which are the basis for quantum operations. The
Rabi frequency $\omega _{R} $ is defined by the amplitude of the
electromagnetic field and usually is much smaller than the energy
difference $\omega_0$ (in frequency units) between the qubit's
states. The ''dressing'' of qubit by the electromagnetic field
splits each level into two giving rise to two new qubits with
energy difference $\omega _{R}$. The spectrum of the multilevel
''qubit + field'' system consists of three lines at the
frequencies $\omega_0$ and $\omega _{0} \pm \omega _{R}$ (the
Mollow triplet~[2]). The second low-frequency electromagnetic
field with the frequency close to the Rabi frequency $\omega _{R}
$ could induce an additional Rabi oscillation on dressed states of
new qubits. These qubits are attracting interest because their
coherence time is longer than that of the initial qubit~ [3 -- 5].
The results of studies of qubits dressed by bichromatic radiation
formed by fields with strongly different frequencies are important
for a wide range of physical objects, including, among others,
nuclear and electron spins, double-well quantum dots, flux and
charge qubits in superconducting systems. In NMR~[6,~7],
EPR~[5,~8,~9] and optical resonance~[10] such investigations are
used in the development of line-narrowing methods.

In this letter, we describe the multiplication of spin qubits at
double resonance in a bichromatic field with strongly different
frequencies. We then show that the operational multiphoton
transitions of dressed qubits can be selected by the choice of
both the rotating frame and the phase of the low-frequency field.
Two important examples of such transitions in the rotating and
doubly rotating frames are presented.

Let an electron spin qubit be in three fields: a microwave (mw)
one directed along the \textit{x} axis of the laboratory frame, a
radio-frequency (rf) one directed along the $z$ axis, and a static
magnetic one also directed along the $z$ axis. The Hamiltonian of
the qubit in these fields can be written as follows:

\begin{equation} \label{GrindEQ__1_}
H=H_{0} +H_{\bot } \left(t\right)+H_{\parallel } \left(t\right).
\end{equation}
Here $H_{0} =\omega _{0} s^{z} $ is the Hamiltonian of the Zeeman
energy of a spin in the static magnetic field $B_{0}$, where
$\omega _{0}=\gamma B_{0}$, and $\gamma $ is the electron
gyromagnetic ratio. Moreover, $H_{\bot } \left(t\right)=2\omega
_{1} \cos \left(\omega t+\phi \right)s^{x}$ and $H_{\parallel }
\left(t\right)=2\omega _{2} \cos \left(\omega _{rf} t+\psi
\right)s^{z}$ are the Hamiltonians of the spin interaction with
linearly polarized mw and rf fields, respectively. $B_{1}$ and
$B_{2}$, $\omega$ and $\omega _{rf}$, and $\varphi$ and $\psi$
denote the respective amplitudes, frequencies, and phases of the
mw and rf fields. Finally, $\omega _{1} =\gamma B_{1}$ and $\omega
_{2} =\gamma B_{2}$ stand for the Rabi frequencies, whereas
$s^{x,y,\, z}$ are the components of the spin operator.

The evolution of the system with the Hamiltonian
\ref{GrindEQ__1_} is described by the Liouville equation for the
density matrix $\rho$:

\begin{equation} \label{GrindEQ__2_}
i\frac{\partial \rho }{\partial t} =[H,\rho ]
\end{equation}
(we set the Planck constant $\hbar =1$). We perform the
transformation ($\rho \to \rho _{1} =U_{1}^{+} \rho U_{1}$, $U_{1}
=e^{-i\omega ts^{z} }$) to the singly rotating frame, which
rotates with frequency $\omega $ around the \textit{z} axis of the
laboratory frame. In this frame, Eq. \ref{GrindEQ__2_} turns
into:

\begin{equation} \label{GrindEQ__3_}
i\frac{\partial \rho _{1} }{\partial t} =[H_{1} ,\rho _{1} ],
\end{equation}
where $H_{1} =U_{1}^{+} HU_{1} =\Delta s^{z} +(\omega _{1}
/2)(s^{+} +s^{-} )+2\omega _{2} \cos (\omega _{rf} t+\psi )s^{z} $
and $\Delta =\omega _{0} -\omega $. The mw phase $\varphi =0$ and
the counter-rotating component of the mw field is neglected. We
also assume that the exact resonance condition is fulfilled
$\Delta =0$, and that $\omega _{1} $, $\omega _{rf} \gg \omega
_{2} $. Upon rotation of the frame around the \textit{y} axis by
the angle of $\pi /2$ ($\rho _{1} \to \rho _{2} =U_{2}^{+} \rho
_{1} U_{2} $, $U_{2} =e^{-i\pi s^{y} /2} $, where $s^{y} =(s^{+}
-s^{-} )/2i$), we obtain:

\begin{equation} \label{GrindEQ__4_}
i\frac{\partial \rho _{2} }{\partial t} =[H_{2} ,\rho _{2} ],
\end{equation}
where $H_{2} =U_{2}^{+} H_{1} U_{2} =\omega _{1} s^{z} -\omega
_{2} \cos (\omega _{rf} t+\psi )(s^{+} +s^{-} )$.

Now, we pass to the interaction representation by choosing the
frame rotating with frequency $\omega _{1} $ around the $z$ axis
($\rho _{2} \to \rho _{3} =U_{3}^{+} \rho _{2} U_{3} $, $U_{3}
=e^{-i\omega _{1} ts^{z} } $). In this frame we have:

\begin{equation} \label{GrindEQ__5_}
i\frac{\partial \rho _{3} }{\partial t} =[H_{3} ,\rho _{3} ],
\end{equation}
where

\noindent $H_{3} =U_{3}^{+} H_{2} U_{3} =-(\omega _{2} /2)s^{+}
\left(e^{i\delta t} e^{-i\psi } +e^{i(2\omega _{rf} +\delta )t}
e^{i\psi } \right)-h.c.$, $\delta =\omega _{1} -\omega _{rf} $,
and $\left|\delta \right|\ll \omega _{1} ,\, \omega _{rf} $ in our
case. Rapidly oscillating ($e^{\pm i2\omega _{rf} t} $) terms in
the Hamiltonian $H_{3} $ can be eliminated by the
Krylov--Bogoliubov--Mitropolsky method~[5,~11,~12]. Averaging over
the period $2\pi /\omega _{rf} $, we obtain the following
effective Hamiltonian up to the second order in $\omega _{2}
/\omega _{rf} $:

\begin{equation} \label{GrindEQ__6_}
H_{3} \to H_{eff} =H_{eff}^{(1)} +H_{eff}^{(2)} .
\end{equation}
In the above equation we have put:

\noindent $H_{eff}^{\ref{GrindEQ__1_}} =\langle H_{3} (t)\rangle
=-(\omega _{2} /2)(s^{+} e^{i\delta t} e^{-i\psi } +h.c.)$,
$H_{eff}^{\ref{GrindEQ__2_}} =\frac{i}{2} \langle [\int
_{}^{t}d\tau  \left(H_{3} (\tau )-\langle H_{3} (\tau )\rangle
\right),\, H_{3} (t)]\rangle =\Delta _{BS} s^{z} $. The symbol
$<...>$ denotes time averaging: $\langle A(t)\rangle
={\frac{1}{T}} \int _{0}^{T}A(t) dt$, where $T=2\pi /\omega _{rf}
$ and $\Delta _{BS} \approx \omega _{2}^{2} /4\omega _{rf} $ is
the Bloch--Siegert-like frequency shift.

After the canonical transformation $\rho _{3} \to \rho _{4}
=U_{4}^{+} \rho _{3} U_{4} $, $U_{4} =e^{-i(\delta t-\psi )s^{z} }
$, the equation

\begin{equation} \label{GrindEQ__7_}
i\frac{\partial \rho _{3} }{\partial t} =[H_{eff} ,\rho _{3} ]
\end{equation}
is transformed into

\begin{equation} \label{GrindEQ__8_}
i\frac{\partial \rho _{4} }{\partial t} =[H_{4} ,\rho _{4} ],
\end{equation}
where $H_{4} =U_{4}^{+} H_{eff} U_{4} =(\delta +\Delta _{BS}
)s^{z} -(\omega _{2} /2)(s^{+} +s^{-} )$.

The diagonalization of the Hamiltonian $H_{4} $ by means of the
rotation operator $U_{5} =e^{-i\xi s^{y} } $ ($\rho _{5} \to \rho
_{5} =U_{5}^{+} \rho _{4} U_{5} $, $H_{5} =U_{5}^{+} H_{4} U_{5}
$) yields:

\begin{equation} \label{GrindEQ__9_}
i\frac{\partial \rho _{5} }{\partial t} =[H_{5} ,\rho _{5} ].
\end{equation}
Here $H_{5} =\varepsilon s^{z} $, $\varepsilon =\sqrt{(\omega _{1}
-\omega _{rf} +\Delta _{BS} )^{2} +\omega _{2}^{2} } $ is the
frequency of the Rabi oscillations between the spin states dressed
simultaneously by the mw and rf field while $\sin \xi =-\omega
_{2} /\varepsilon $, $\cos \xi =(\omega _{1} -\omega _{rf} +\Delta
_{BS} )/\varepsilon $.

By using Eqs. \ref{GrindEQ__2_} -- \ref{GrindEQ__9_}, the
density matrix in the laboratory frame (LF) can be written as:

\begin{equation} \label{GrindEQ__10_}
\rho (t)=U_{1} U_{2} U_{3} U_{4} U_{5} e^{-iH_{5} t} \rho _{5}
(0)e^{iH_{5} t} U_{5}^{+} U_{4}^{+} U_{3}^{+} U_{2}^{+} U_{1}^{+}
,
\end{equation}
where
\begin{equation} \label{GrindEQ__11_}
\rho _{5} (0)=U_{5}^{+} U_{4}^{+} (0)U_{3}^{+} (0)U_{2}^{+}
U_{1}^{+} (0)\rho (0)U_{5} U_{4} (0)U_{3} (0)U_{2} U_{1} (0),
\end{equation}
and $U_{1} (0)=1$, $U_{3} (0)=1$, $U_{4} (0)=e^{-i\psi s^{z} } $.

Initially, the qubit is in the ground state and $\rho
(0)=1/2-s^{z} $. The absorption signal $\upsilon (t)=Sp\{ \rho
(t)(s^{+} -s^{-} )\} /2i$ in the laboratory frame can be derived
from Eqs. \ref{GrindEQ__10_} and \ref{GrindEQ__11_}:

\[\upsilon (t)=\left(\langle 1|\rho (t)|2\rangle -\langle 2|\rho (t)|1\rangle \right)/2i=\]
\[=(1/2)\sin \xi \cos \xi \cos \psi \sin \omega t+(1/16)\{ 4\sin \xi \sin \psi [\cos (\omega +\varepsilon )t-\cos (\omega -\varepsilon )t]+\]
\[-4\sin \xi \cos \xi \cos \psi [\sin (\omega -\varepsilon )t+\sin (\omega +\varepsilon )t]+2\sin ^{2} \xi \sin 2\psi [\cos (\omega +\omega _{rf} )t+\cos (\omega -\omega _{rf} )t]+\]
\[+\left(\left(\cos \xi -1\right)^{2} +\left(\cos ^{2} \xi -1\right)\cos 2\psi\right)\times[\sin (\omega +\omega _{rf} -\varepsilon )t-\sin (\omega -\omega _{rf} +\varepsilon )t]+\]
\[+\left(\cos ^{2} \xi -1\right)\sin 2\psi \times[\cos (\omega -\omega _{rf} +\varepsilon )t+\cos (\omega +\omega _{rf} -\varepsilon )t]+\]
\[+\left(\left(\cos \xi +1\right)^{2} +\left(\cos ^{2} \xi -1\right)\cos 2\psi \right)\times[\sin (\omega +\omega _{rf} +\varepsilon )t-\sin (\omega -\omega _{rf} -\varepsilon )t]+\]
\begin{equation} \label{GrindEQ__12_}
+\left(\cos ^{2} \xi -1\right)\sin 2\psi\times[\cos (\omega -\omega _{rf} -\varepsilon )t+\cos (\omega
+\omega _{rf} +\varepsilon )t]\}.
\end{equation}

\begin{figure}[]
    \centering
    \vspace{-1 cm}
    \includegraphics{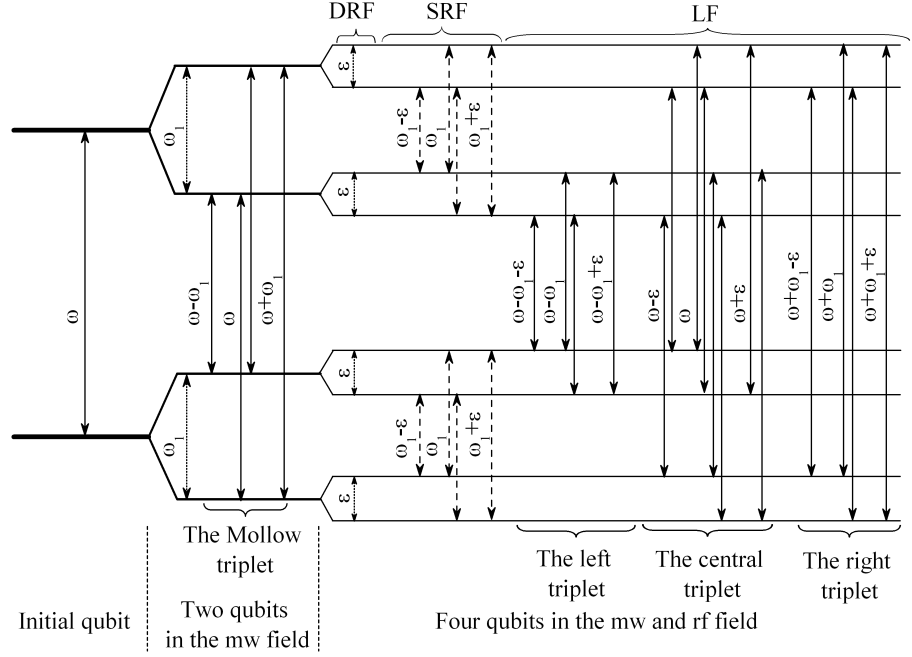}
    \vspace{-1 cm}
    \caption{Energy-level diagram of a qubit and
transitions created by a bichromatic field at double resonance
($\omega _{0} =\omega $, $\omega _{1} =\omega _{rf} $).}
\end{figure}

The resonant interaction between the mw field and the qubit
creates its dressed states and two new qubits with energy
splitting equal to the Rabi frequency $\omega _{1} $, as shown in
Fig.1. The rf field with the frequency $\omega _{rf} $, which is
close to the Rabi frequency $\omega _{1} $ of the new qubits,
''dresses'' these qubits, giving rise to four qubits with the
energy splitting $\epsilon$. Allowed transitions between states of
these qubits afford nine spectral lines observed in the laboratory
frame.

Figs. 2 and 3 show the time evolution of absorption signals and
their Fourier spectra of dressed qubits under conditions typical
for EPR.

\begin{tabular}{p{3.2in}p{3.2in}}
{
\includegraphics[width=80 mm]{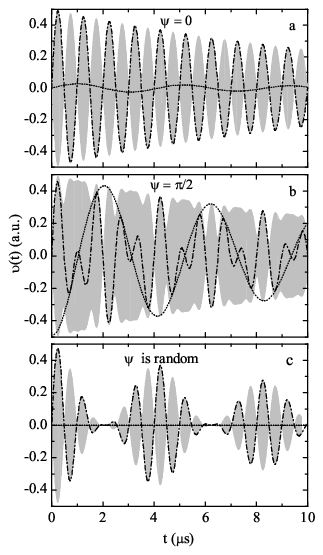}
}
&
{
\includegraphics[width=80 mm]{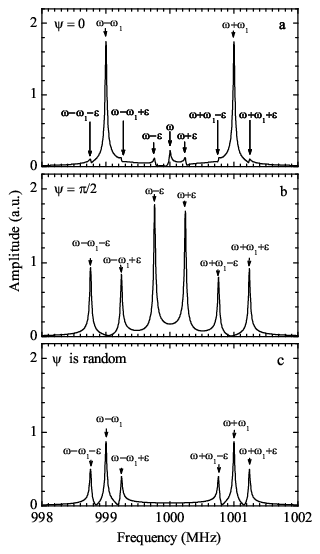}
}
\\
 {\footnotesize Figure 2: Time evolution of the absorption signals in the laboratory
(solid line), singly rotating (dot line) and doubly rotating (dash
line) frames. The signals were obtained for the following
parameters of the bichromatic field: $\omega =\omega _{0} $,
$\omega _{1} =\omega _{rf} =2\pi $1.0 MHz, $\omega _{2} =2\pi
$0.24 MHz, using the exponential decay function with \textit{T} =
16 $\mu$s.}& {
\footnotesize Figure 3: Fourier spectra of the absorption signals in the
laboratory frame shown in Fig. 2 by a solid line.}
\end{tabular}

For the rf phase $\psi =0$, three triplets with the intensive
central lines at $\omega -\omega _{1} $, $\omega $ and $\omega
+\omega _{1} $ are formed (Fig. 3a). The less intensive sidebands
have the frequencies $\pm \varepsilon $ relative to each of the
central lines. For the rf phase $\psi =\pi /2$, the central lines
in the triplets vanish (Fig. 3b), each triplet turning into an
doublet. When the rf phase is random, averaging over a
sufficiently large number of experiments (at the uniform
distribution of the phase in the interval from 0 to 2$\pi$) leads
to a complete removal of the central triplet. The differences of
line's intensities in the two residual triplets ($\omega -\omega
_{1} $, $\omega -\omega _{1} \pm \varepsilon $ and $\omega +\omega
_{1} $, $\omega +\omega _{1} \pm \varepsilon $) become smaller
(Fig. 3c).

There is the possibility of selecting the observed transitions of
four qubits by employing the rotating frame. In the singly
rotating frame (SRF), the absorption signal described by the
density matrix $\rho _{1} (t)$ (Eq. (4)) can by written as

\begin{equation} \label{GrindEQ__13_}
\begin{array}{l}
{\upsilon _{1} \left(t\right)=(1/8)\left[2\sin ^{2} \xi \left(\sin \omega _{rf} t+\sin \left(\omega _{rf} t+2\psi \right)\right)+\right. }\\
{+\left(1+\cos \xi \right)^{2} \sin \left(\omega _{rf}
+\varepsilon \right)t-}\\{ -\left(1-\cos ^{2} \xi \right)\sin
\left(\left(\omega _{rf} +\varepsilon \right)t+2\psi \right)+} \\
{\left. +\left(1-\cos \xi \right)^{2} \sin \left(\omega _{rf}
-\varepsilon \right)t-\right.}
\\{\left. -\left(1-\cos ^{2} \xi \right)\sin
\left(\left(\omega _{rf} -\varepsilon \right)t+2\psi
\right)\right].}
\end{array}
\end{equation}

Fig. 4 shows the Fourier spectra of signals given by Eq.
\ref{GrindEQ__13_} under the same conditions as in Figs. 2 and
3. For the random rf phase, the absorption signal has three
comparable oscillating components with frequencies $\omega _{1} $
and $\omega _{1} \pm \sqrt{\omega _{2}^{2} +\Delta _{BS}^{2} } $
(Fig. 4c). For the rf phase $\psi =0$, the sidebands are smaller
than those at the random rf phase by the factor $\Delta _{BS}
/\sqrt{\omega _{2}^{2} +\Delta _{BS}^{2} } $ (Fig. 4a). When we
use $\psi =\pi /2$, the component with frequency $\omega _{1} $
vanishes and the sidebands are comparable to those at the random
rf phase (Fig. 4b). Note that the high-frequency sideband is
always more intensive than the low-frequency one.
\begin{center}
\begin{tabular}{p{3.2in}}
{
\includegraphics[width=80 mm]{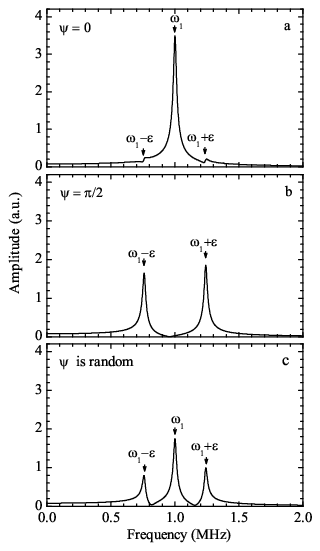}
}

\\
 {
\footnotesize Figure 4: Fourier spectra of the absorption signals in the singly
rotating frame shown in Fig. 2 by a dot line.}
\end{tabular}
\end{center}
Upon the rotating wave approximation ($\Delta _{BS} \to 0$), it
follows from Eq. \ref{GrindEQ__13_} that for $\psi =0$ only the
component with frequency $\omega _{1} $ remains. At the same time,
for both $\psi =\pi /2$ and the random rf phase, the intensities
of the sidebands are equal. The equalization of sidebands can be
used to indicate the validity of the rotating wave approximation.
On the contrary, their asymmetry reveals the effect of the
counter-rotating component of the rf field. Such asymmetry was
observed in the dressed Rabi oscillations using the EPR experiment
with random rf phase [5].

Note that, upon the resonant monochromatic interaction, the Mollow
triplet is formed by the transitions between the dressed states of
the ground and excited levels of the initial qubit. Similarly, at
the doubly resonant bichromatic interaction, the triplet in the
singly rotating frame is formed by the transitions between the
split states of both the ground and excited levels of the initial
qubit (Fig. 1).

We now provide the expression for the absorption signal in the frame
described by the density matrix $\rho _{5} (t)=e^{-iH_{5} t} \rho
_{5} (0)e^{iH_{5} t} $, where $\rho _{5} (0)$ is given by Eq.
\ref{GrindEQ__11_}. In this doubly rotating frame (DRF), the
absorption signal can be written as follows:
\begin{equation} \label{GrindEQ__14_}
\upsilon _{5} (t)=\left(\langle 1|\rho _{5} (t)|2\rangle -\langle
2|\rho _{5} (t)|1\rangle \right)/2i=\left(\cos \xi \cos \psi \sin \varepsilon t-\sin \psi \cos
\varepsilon t\right)/2.
\end{equation}

According to Eq. \ref{GrindEQ__14_}, the absorption signal in
the doubly rotating frame is caused by the transitions between
spin states dressed simultaneously by the mw and rf fields. At the
exact resonance ($\omega _{1} =\omega _{rf} $), the signal for
$\psi =0$ is smaller than the signal for $\psi =\pi /2$ by the
factor $\Delta _{BS} /\sqrt{\omega _{2}^{2} +\Delta _{BS}^{2} } $.
If $\Delta _{BS} \to 0$, the signal for $\psi =0$ disappears. In
this case, for $\psi =\pi /2$, the absorption signal oscillates
with the Rabi frequency $\omega _{2} $. So, for $\psi =0$, the
absorption signal $\upsilon _{5} $ is fully due to the
counter-rotating component of the rf field and its amplitude is
proportional to the value of the Bloch--Siegert shift $\Delta
_{BS} $.

In conclusion, we have studied the evolution of spin qubits at the
double resonance ($\omega _{0} =\omega $, $\omega _{1} =\omega
_{rf} $) with a bichromatic field, consisting of transverse
(high-frequency) and longitudinal (low-frequency) components. We
have found that the double ''dressing'' of an initial qubit by the
bichromatic field forms four new qubits with a smaller energy
splitting, giving rise to multiphoton transitions. In the
laboratory frame, three triplets correspond to the transitions
between states of these qubits. The transition amplitudes depend
strongly on the phase of the low-frequency field. The
counter-rotating component of the low-frequency field causes the
asymmetry of sidebands in the triplets. After taking into account
this component, the errors in operations with qubits on dressed
states in the strong-field regime are minimized. The types of
operational multiphoton transitions can be selected by the choice
of the rotating frame: one triplet, ($\omega _{1} $, $\omega _{1}
\pm \varepsilon $), can be observed in the singly rotating frame,
and only the transition at the frequency $\varepsilon $ is
realized in the doubly rotating frame.

\vfill\eject

\end{document}